\documentclass[useAMS,usenatbib]{mnras}

\usepackage[utf8]{inputenc}
\usepackage{amssymb,amsmath}
\usepackage{graphicx}
\usepackage[dvipsnames]{xcolor}
\hypersetup{colorlinks=true,allcolors=violet}
\usepackage{todonotes}
\usepackage{academicons}
\usepackage{xcolor}

\newcommand{\be}{\begin{equation}}
\newcommand{\ee}{\end{equation}}
\def\bear#1\ear{\begin{align}#1\end{align}}

\renewcommand{\mathbf}[1]{\mbox{\boldmath $#1$}}

\title[Stellar Mid-life Crisis and Subcritical Dynamos]{Stellar Mid-life Crisis: Subcritical Magnetic Dynamos of Solar-like Stars and the Break-down of Gyrochronology}
\author[Tripathi, Nandy \& Banerjee]{
Bindesh Tripathi$^{1,2,3}$ ,
Dibyendu Nandy$^{1,4}$\thanks{dnandi@iiserkol.ac.in},
Soumitro Banerjee$^{4}$
\\
$^{1}$ Center of Excellence in Space Sciences India, Indian Institute of Science Education and Research Kolkata, Mohanpur 741246, India\\
$^{2}$ University of Wisconsin-Madison, Madison, Wisconsin 53706, USA\\
$^{3}$ St. Xavier's College, Tribhuvan University, Kathmandu, Nepal\\
$^{4}$ Department of Physical Sciences, Indian Institute of Science Education and Research Kolkata, Mohanpur 741246, India
}
\begin{document} 
 
\date{} 

\maketitle

\begin{abstract}
Recent observations reveal the surprising breakdown of stellar gyrochronology relations at about the Sun's age hinting that middle-aged, solar-like stars transition to a magnetically inactive future. We provide a theoretical basis for these intriguing observations inspired by simulations with a mathematical dynamo model that can explore long-term solar cycle fluctuations. We reproduce the observed bimodal distribution of sunspot numbers, but only for subcritical dynamos. Based on a bifurcation analysis, we argue that ageing of solar-like stars makes the magnetically-weak dynamo regime readily accessible. Weak magnetic field production in this regime compromises wind-driven angular momentum losses thus disrupting the hegemony of magnetic braking on stellar rotational spin-down. This hypothesis of {\emph{subcritical magnetic dynamos}} of solar-like stars provides a self-consistent, unifying physical basis for a diversity of solar-stellar phenomena such as why stars beyond their mid-life do not spin-down as fast as in their youth, the break-down of stellar gyrochronology relations, the observed bimodal distribution of long-term sunspot observations and recent findings suggesting that the Sun may be transitioning to a magnetically inactive future. 
\end{abstract}

\section{Introduction} 
The widely used diagnostic of stellar age rotation relationship that is the basis of gyrochronology has recently been found to be unreliable after stellar midlife \citep{vansaders2016}. The standard understanding of angular momentum loss mediated braking of rotation due to magnetically-coupled stellar wind cannot explain this. It has been reported that the Sun is currently going through a weakened braking phase at about its midlife of 4.5 billion years \citep{vansaders2016}. What causes this weakened magnetic braking is an intriguing question. Latest measurements confirm that the Sun is less magnetically active than similar stars \citep{reinhold2020}. This has spurred debates and controversy on whether the Sun is already transitioning to a magnetically inactive future \citep{metcalfe2020, reinholdreplies2020}; studies exploring rotation period magnetic cycle period relationships support this hypothesis \citep{metcalfe2017, metcalfe2016}. Since stellar magnetic activity is ultimately a product of the magnetohydrodynamic (MHD) dynamo mechanism, a causal basis could exist between the nature and evolution of stellar dynamo action and these observations. Inferring the nature of magnetic dynamos in Solar-like stars at about their midlife, therefore, could be the Rosetta stone to explain these apparent disparate observations.

With an aim to characterize solar-like dynamos at around their midlife -- i.e., the current age of the Sun we look at long-term solar observations. These reveal intriguing phases of reduced sunspot activity that are termed as grand minima episodes, the most recent being the Maunder minimum between 1645 to 1715 AD \citep{eddy_maunder_1976}. Several long-term solar activity reconstructions based on cosmogenic isotopes like ${Be}^{10}$ in ice cores and $C^{14}$ in tree rings suggest that such grand minima have also occurred in the past \citep{usoskin_grand_2007}; reconstructed data reveal 20 grand minima in the last 9000 years \citep{usoskin2016solar}. Such grand minimum have recently been confirmed as a special mode of solar dynamo operation, distinct from the regular activity cycle mode \citep{usoskin_evidence_2014}.

Similar low activity phases have been observed in other Sun-like stars, which have been dubbed ``Maunder-minima stars'' \citep{baliunas1995chromospheric}; the nature of magnetic dynamos in these stars is hotly debated. A crucial observation is that the main sequence age of these low activity ``Maunder-minima stars'' are typically similar or older compared to the Sun. This is thought to be due to stellar spin down through steady angular momentum losses mediated via stellar winds and consequent reduction in the efficiency of the dynamo mechanism \citep{nandy2004,brun2015}. It has been speculated that this might be related to a drastic change in the nature of solar-stellar dynamos at the current solar age \citep{metcalfe2019,metcalfe2017,metcalfe2016}.

Are the wandering of the Sun's magnetic activity between the two distinct modes of low and regular activity cycles, the occurrence of ``Maunder minima stars'' and the observed break-down of the rotation rate--age relationship connected? Based on a reduced dynamo model that imbibes the spiritual essence of Babcock-Leighton and turbulent dynamos, and utilizing stochastically forced, non-linear delay differential equations, we show these diverse phenomena can be explained on the basis of bimodality of the underlying dynamo mechanism. Based on this analysis, we argue that a stellar evolution mediated  shift in the dynamo efficiency enables a transition across supercritical to subcritical states that makes the full dynamical regime -- including very weak activity states -- accessible to an aging star.

The reconstructed sunspot number suggests that the occurrence of grand minima is not a cyclic phenomenon but an outcome of an underlying generation mechanism that is inherently stochastic and/or chaotic \citep{usoskin_grand_2007}. In a chaotic dynamo, the long time-scale `supermodulation' could be related to symmetry changes in magnetic fields \citep{weiss2016}. The current understanding of large-scale magnetic field generation in Sun-like stars is based on the recycling of toroidal and poloidal magnetic field components \citep{parker1955hydromagnetic,charbonneaureview}. The toroidal field is generated from the poloidal field due to stretching by differential rotation. For the current Sun, observations and data-driven model reconstructions of solar activity \citep{dasi_espuig2010, cameron2015,prantika2018} demonstrate that the dominant mechanism for poloidal field creation is the so-called Babcock-Leighton (BL) mechanism, i.e., the decay and dispersal of bipolar sunspot pairs which contribute to a large-scale dipolar field \citep{babcock1961,leighton1969}. This mechanism relies on the existence of strong toroidal flux tubes in the Sun's convection zone, which is not the case during grand minima episodes. It has been demonstrated recently that a mean-field dynamo $\alpha$-effect capable of working on weaker fields can ``rescue'' the solar cycle from a grand minimum \citep{passos2014,hazra_stochastically_2014} and therefore, must be operational in stellar convection zones. Direct numerical simulations utilizing the full magnetohydrodynamic (MHD) system of equations or spatially extended mean-field dynamo models are too expensive computationally for the necessary long-term simulations to explore the range of possible dynamical solutions a solar-like dynamo mechanism may exhibit over long-term. For our numerical bifurcation analysis, we adapt a reduced BL dynamo model based on delay differential equations that has been found very useful in providing insights on long-term solar cycle behavior \citep{wilmot-smith_time_2006, hazra_stochastically_2014}.

In Babcock-Leighton dynamos, the sources of the toroidal and the poloidal fields are spatially segregated as they act across a distributed domain in the solar convection zone (SCZ). Such spatial segregation necessitates a finite time for flux transport across the SCZ, introducing time delays in the communication between the two source layers \citep{jouve2010a}. These time delays have been shown to create a dynamical memory even in stochastically forced dynamo systems which allows a predictive window for the activity \citep{yeates_exploring_2008,karak_nandy2012, prantika2018}.

The evolution of the toroidal field ($B_\phi$) and the vector potential ($A$) for the poloidal field in the reduced BL dynamo model are governed by the following delay differential equations: 
\begin{eqnarray}
&&\frac{dB_\phi (t)}{dt} = \frac{\omega }{L} A(t-T_0)-\frac{B_\phi (t)}{\tau }\label{eqn1}\\
&&\frac{dA(t)}{dt} = \alpha _{BL} \cdot {f} \left(B_\phi \left(t-T_1\right)\right) \cdot B_\phi (t-T_1) -\frac{A(t)}{\tau } \label{eqn2}
\end{eqnarray}	
where $T_0$ and $T_1$ are the time delays in the conversion of the poloidal flux to the toroidal one and vice-versa.
The interplay amongst meridional circulation, turbulent diffusion and turbulent pumping mediated flux transport \citep{hazra2016} determine $T_0$. While there have been suggestions that meridional flow variations can explain the observed stellar activity amplitude-period relationship \citep{nandy2004}, numerical simulations exploring meridional flow dependence on stellar rotation rates indicate the necessity of more complex flow profiles or additional mechanisms to explain this \citep{jouve2010b}. Thus we reiterate that $T_0$ must be interpreted as a generic flux transport timescale, not necessarily determined by stellar meridional flows alone. The turbulent dissipation timescale of the field is $\tau$, $\omega$ is the difference in rotation rates across the SCZ of length $L$, and $\alpha _{BL}$ represents the BL source term for the poloidal field. The quenching function $f$ constrains the poloidal source to work within a finite range of toroidal field strength with a non-zero lower bound. This is motivated from the perspective that weak toroidal flux tubes are shredded by turbulence and cannot form sunspots and very strong flux tubes are immune to the Coriolis force, which causes them to emerge at the solar surface without any tilt. This effectively quenches the BL source (for further details and motivation, see \citep{wilmot-smith_time_2006,hazra_stochastically_2014}). The efficiency of the dynamo is measured by a non-dimensional number called the dynamo number, which is defined as $N_D = \omega \alpha_{BL} \tau^2/L$.
 
\section{Results and Discussion \label{sec:results}}

\begin{figure*}
\centering 
\hspace{-0.75cm}\includegraphics[height = 0.33\textwidth]{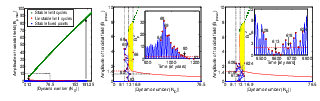} \vspace{-0.83cm}
\caption{\label{bifn diagram} (color online). (Left panel) Bifurcation diagram of toroidal field's amplitude. The middle red curve (unstable limit cycles) between the upward and the downward arrows separates the basins of attraction for the oscillating solutions (stable limit cycles) and the decaying solutions (stable fixed points). The dotted rectangular region at the bottom left is zoomed in and shown in the middle panel. (Middle panel) The region within the dotted vertical lines at $|N_D|=9.1$ and $|N_D|=16.9$ is the dynamo working region with 30\% fluctuation in the dynamo number $|{N_{\mathrm{D}}}^{\mathrm{mean}}|=13$. The dashed-dotted and dashed horizontal lines are the sunspot formation threshold and the dynamo operating threshold (below which the Babcock-Leighton (BL) process alone cannot recover the decaying cycles). The yellowish shaded region represents the basin of attraction for the oscillatory solutions. The numbers (66, 68, 69, 80, 83) are cycle numbers counted from the beginning of the simulation. The fluctuating dynamo numbers (depicted by black filled circles) are noted at each cycle maximum of the time series (shown in the inset). (Right panel) Typical onset of (and exit from) a grand minimum phase when an additive magnetic noise is employed with a typical root-mean-squared (rms) value of about 17\% of the rms value of the BL source in equation (\ref{eq3}). The parameters used are $\omega/L = -0.34$, $\tau = 15$, $T_0 = 2$, $T_1 = 0.5$ yrs.}
\end{figure*}

Long-term simulations with this dynamo model shows hysteresis (see Fig. \ref{bifn diagram} left panel).  There is a hint of similar behavior in earlier studies with non-linear mean field dynamos and $3$-dimensional MHD simulations of turbulent dynamos \citep{kitchatinov_dynamo_2010, karak_hysteresis_2015, dalton2020}. However, these computationally expensive numerical simulations precluded an adequate analysis of the distribution function of a large number of magnetic cycles. The results from our model (Fig. \ref{bifn diagram} left panel) substantiate the proposition and demonstrate that hysteresis arises from two distinct modes of dynamo activity. Note that the results are qualitatively robust with respect to choice of parameters within the regimes established in \citet{hazra_stochastically_2014} and \citet{wilmot-smith_time_2006}. 

The dynamo cannot self-consistently recover from a grand minimum once it is nudged into such a phase by stochastic fluctuations in the BL source term (Fig. \ref{bifn diagram} middle panel). Our extensive parameter space study shows that the BL dynamo shut-down threshold (dashed  horizontal line in the inset of the middle panel of Fig. \ref{bifn diagram}) shifts upward to engulf more cycles into grand minima episodes when the timescale $T_0$ is increased in the diffusion-dominated regime (or decreased in the advection-dominated regime) as defined in \citet{wilmot-smith_time_2006}.
 
We find that the addition of a ``magnetic noise'' allows the dynamo to recover from grand minima episodes as well as sustains hysteresis. Incorporation of additive noise, $\epsilon (t)$, modifies equation (\ref{eqn2}) to the form:
\begin{eqnarray} \label{eq3}
\hspace{-0.02cm}\frac{dA(t)}{dt} = \alpha_{BL} \cdot {f} \left(B_\phi \left(t-T_1\right)\right) \cdot B_\phi (t-T_1) -\frac{A(t)}{\tau } + \epsilon (t)
\end{eqnarray}
where $\epsilon (t)$ is uniform and white in time \citep{charbonneau_multiperiodicity_2001,charbonneau_intermittency_2004,charbonneau2007fluctuations} and has zero mean. Since the noise has zero mean, it sometimes builds up the poloidal field while at other times, reduces the existing dipolar field of the Sun. Magnetic noise definitely abounds in the solar convection zone, especially near the sub-surface layers of the Sun where the convection is much more turbulent. Small-scale dynamo action in this region \citep{meneguzzi1989turbulent,nordlund1992dynamo} produces magnetic field that is highly uncorrelated in space and time. Such noise continuously regenerates and replenishes the magnetic field in the upper layers of the Sun \citep{cattaneo_interaction_2003,hagenaar2003properties,charbonneau_intermittency_2004} and our results (Fig. \ref{bifn diagram} right panel) indicate this could be enough to recover solar-stellar dynamos from a grand minima over a growth time that is dependent on the level of imposed noise. 

\begin{figure}
\hspace{-0.65cm}\includegraphics[width=0.52\textwidth]{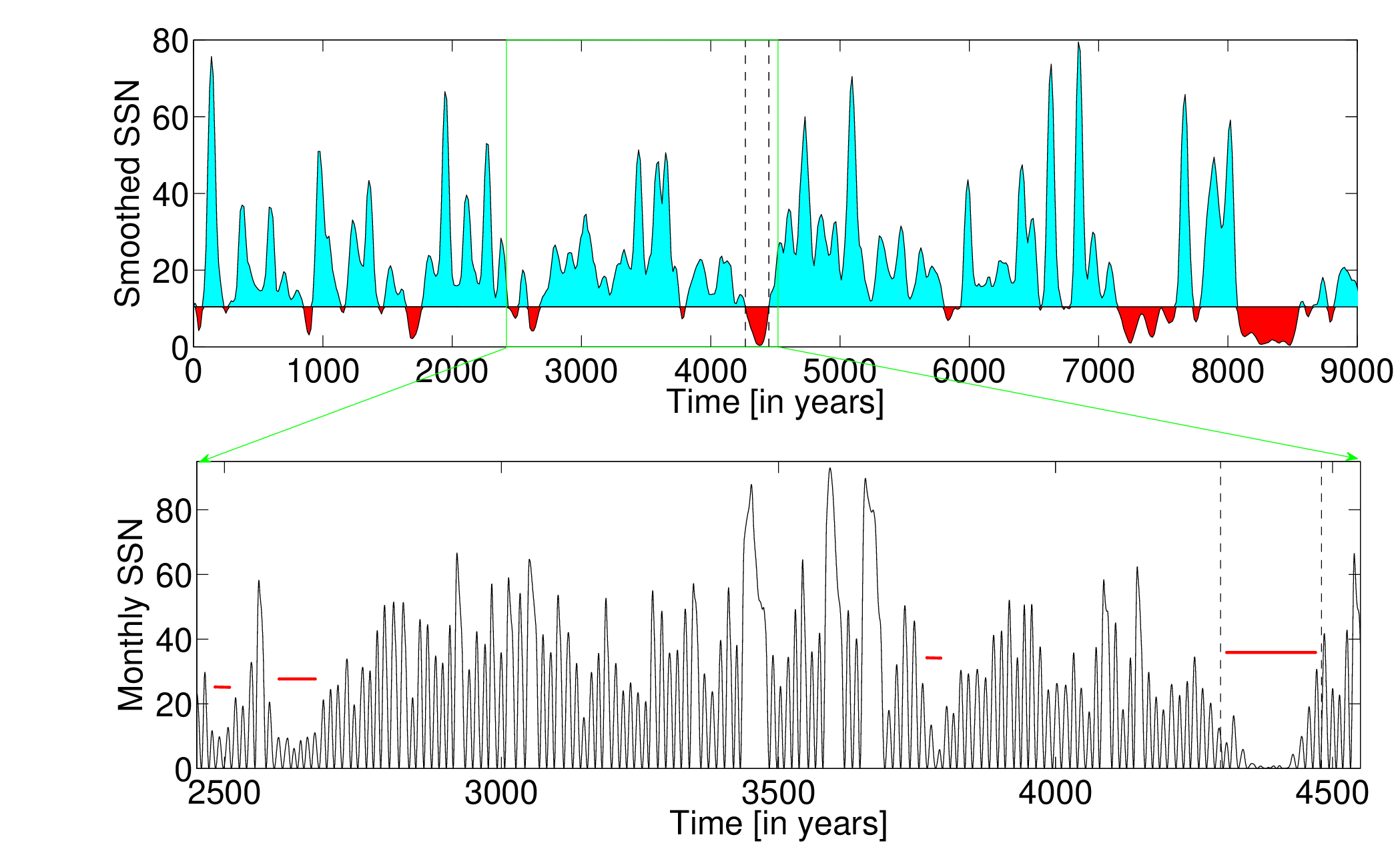}
\caption{\label{smoothSSN} (color online). (Upper panel) Temporal variation of the smoothed SSN from a 9000 yr simulation. The shaded regions below the horizontal line represent grand minima episodes. (Lower panel) Variation of monthly SSN for a selected duration of 2100 yrs. The monthly SSN is assumed to be a measure of the toroidal field energy (i.e., $SSN = B_\phi^{2}$). The (broken) horizontal lines depict the extent of different grand minima epochs. The parameter used is $|{N_{\mathrm
{D}}}^{\mathrm{mean}}|=13$ and the level of fluctuations in $|N_{\mathrm{D}}|$ is $30\%$. The rms value of the additive noise imposed here is about 24\% of the rms value of the Babcock-Leighton source. Other parameters of the model are set at $\omega/L = -0.34$, $\tau = 15$, $T_0 = 2$, $T_1 = 0.5$ yrs.}
\end{figure}

A time series of the smoothed cycle-averaged sunspot number from an additive-noise incorporated Babcock-Leighton dynamo model is shown in Fig. \ref{smoothSSN}, upper panel. The smoothing has been done using the same procedure used by \citet{usoskin2016solar}, i.e., we first average the monthly sunspot number (SSN) over the usual cycle period and apply the Gleissberg low pass 1-2-2-2-1 filter. The shaded regions below the horizontal line signify grand minima, which are defined as epochs during which the smoothed SSN falls below 50\% of its average value for at least 3 consecutive decades. We observe 13 to 18 grand minima (18 in Fig. \ref{smoothSSN}) in different realizations of stochastic fluctuations in a typical 9000 yr simulation, which is in close agreement with the reconstructed data identifying 20 grand minima in the last 9 millennia; this is of course sensitive to model parameter choices \citep{choudhuri2012}. 
 
\begin{figure} 
\centering
\hspace{-0.9cm}\includegraphics[width=0.5\textwidth]{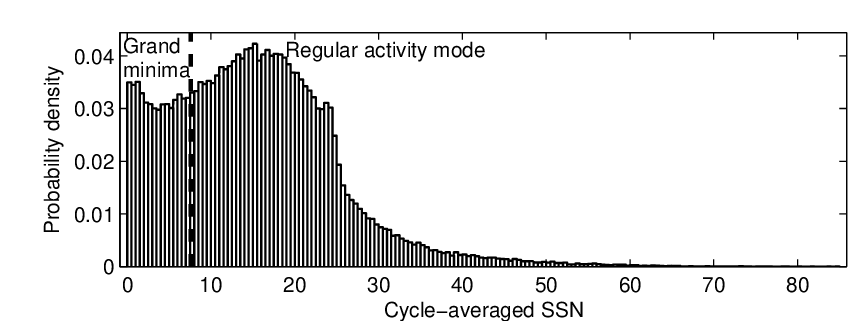}
\caption{\label{Bimodal_distribution} Bimodal distribution of probability density of toroidal magnetic energy obtained from the reduced dynamo model, which is gleaned from a 2 million years simulation run. The relevant dynamo parameters are $|{N_{\mathrm{D}}}^{\mathrm{mean}}|=13$, level of fluctuations in $|N_{\mathrm{D}}|$ is $20\%$ and the rms value of the additive noise is about $17\%$ of the rms value of the Babcock-Leighton source. Other parameters are fixed at $\omega/L = -0.34$, $\tau = 15$, $T_0 = 2$, $T_1 = 0.5$ yrs.}
\end{figure}

\begin{figure}
\centering \vspace{-0.5cm}
\hspace{-0.8cm}\includegraphics[width=0.5\textwidth]{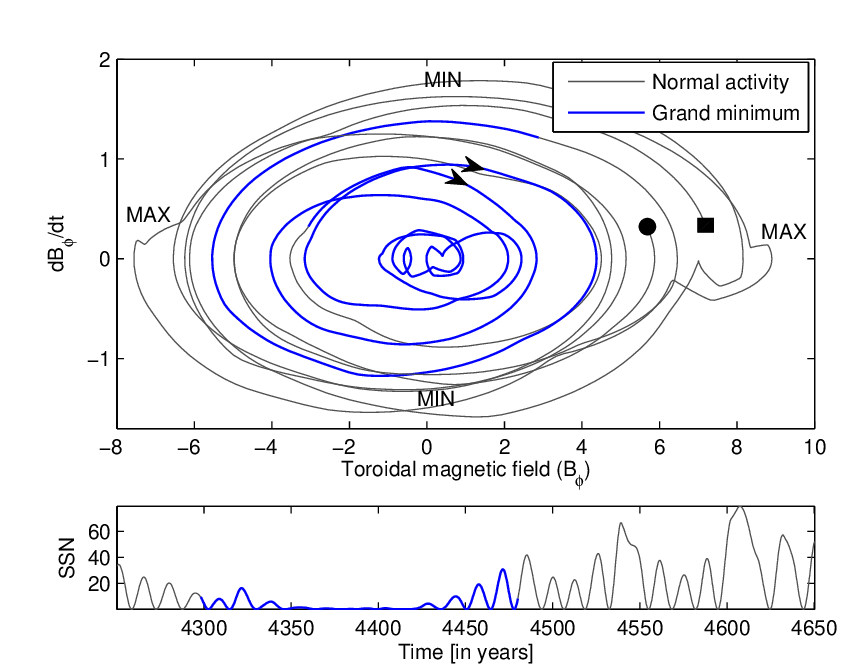}
\caption{\label{Phase space collapse} (color online). (Upper panel) Phase space collapse seen in a typical 400 year simulation when a grand minimum phase was observed. This is the same grand minimum shown in Fig. \ref{smoothSSN} (lower panel) within the dotted vertical lines at around time = 4400 year. The filled circle (square) represent the beginning (end) of the considered 400 yrs. (Lower panel) Corresponding temporal variation of sunspot number in the simulation.}
\end{figure}

What is fascinating is that the assimilation of magnetic noise in our model generates magnetic cycles that qualitatively reproduce the observed bimodal distribution of toroidal field energy; see Fig. \ref{Bimodal_distribution} and for a comparison with observations, Fig. 3 in \citet{usoskin_evidence_2014}. This implies that grand minima episodes cannot be considered as events that correspond to a tail of a single regular mode of solar activity, but it is in itself a separate state of dynamo operation. Most previous studies of grand minima relied on fluctuations in a single regular mode of activity cycles \citep{olemskoy2013,moss2008,brandenburg2008modeling} and thus could not explain the observed bimodal distribution. This bimodal nature in the probability density of cycle-averaged SSN is, in fact, due to the hysteresis present in the model \citep{kitchatinov_parametric_2015}. Hysteresis in a certain parameter regime emerges with the existence of bistable solutions, which emanate in our model due to the non-zero lower operating threshold of the Babcock-Leighton process and the shorter poloidal-to-toroidal flux conversion timescale. Numerical experiments with our dynamo model show that if we remove the lower operating threshold and increase the time delays simultaneously, hysteresis and the bimodal nature of the probability density of cycle-averaged sunspot number is no longer manifested.

Our model also qualitatively reproduces the toroidal field's phase space collapse; compare Fig. \ref{Phase space collapse} upper panel here with fig. 10 in \citet{lopes2014}, which \citet{passos2011grand} unsuccessfully tried to achieve using their low dimensional model with stochastic fluctuations in $\alpha$-effect alone. The collapse in the phase space implies that a grand minimum episode results in strong intermittency in magnetic cycles and that weak intermittency (or moderate amplitude cycles) cannot be attributed to a grand minimum epoch. It is interesting to note in Fig. \ref{Phase space collapse} (upper panel) that the polarity reversal of magnetic cycles, albeit far less regular, still occurs during the grand minimum when the Babcock-Leighton source is dysfunctional. This is because the stochastically injected poloidal field due to magnetic noise still provides a seed for differential rotation to generate toroidal fields. A combination of statistical fluctuations in the right direction can then nudge the weak toroidal field such that it exceeds the lower operating threshold and revive the large-scale BL dynamo that produces sunspots. The lower panel of Fig. \ref{Phase space collapse} portrays the corresponding temporal variation of toroidal field energy depicting this dynamics. 

We observe the bimodal distribution of sunspot numbers only in the (significantly) subcritical regime of the dynamo. Earlier works in this context were inconclusive and conflicting \citep{karak_hysteresis_2015,kitchatinov2017}. The dynamo hysteresis in our model sustains coexisting oscillating and decaying solutions for similar dynamo parameters, which is not possible for supercritical dynamos.  Thus, our theoretical analysis establishes that the observed bimodal distribution of decadal sunspot numbers is a direct manifestation of the subcritical nature of the current solar dynamo (at midlife of its evolution as a star). 

Over its life time, a Sun-like star spins down due to angular momentum loss through magnetically and thermally driven stellar winds. Aging and lower rotation rates (i.e., higher rotation period $P_{\Omega}$) increases the stellar Rossby number (${R_o}$ = $P_{\Omega}/\tau_c$, where $\tau_c$ is the convective turn-over time) -- a parameter used widely in observational studies of stellar activity. The efficiency of the dynamo generation of magnetic fields represented by the dynamo number scales as ${N_\mathrm{D}} \sim 1/{R_o}^2$. Based on our earlier analysis we postulate that the aging of a star -- with a concomitant increase in ${R_o}$ and decrease in ${N_\mathrm{D}}$ -- makes the subcritical dynamo regime with bimodal activity distribution accessible around the midlife of its evolution on the Main Sequence. 

\section{Conclusion} \label{sec:conclusion}

Our analysis is relevant only for stars which continue to exhibit solar-like rotation and thus not applicable for anti-solar type differential rotation, see, e.g., \cite{karak2015a,karak2020,brandenburg2018}. A systemic change in global coronal magnetic field configuration (e.g., from closed to open loops) may also impact wind properties and the nature of magnetic braking which we do not address here.

Our findings have several important consequences. We establish that Maunder-like grand minima are a distinct mode of dynamo operation possible only in sub-critical dynamo regimes. Stellar observations indicate the existence of the low activity ``Maunder minima stars'', which are similar or older than the Sun \citep{baliunas1995chromospheric} and have high ${R_o}$. Our framework suggests these stars are in the sub-critical regime, where accessing the low activity mode of the bimodal stellar dynamo becomes possible. Based on our analysis, we postulate that this is not due to a drastic change in the underlying dynamo mechanism itself, but merely the opening up of the possibility of existing in the low activity mode -- wherein, very low dynamo output drastically reduces magnetically driven stellar winds and angular momentum loss rates. At about stellar mid-life, the magnetically-weak sub-critical states are readily attainable. These weak-magnetic epochs disrupt the hegemony of magnetic braking on stellar rotation. This argument is consistent with the suggestion by \citet{metcalfe_tess_2020} -- namely, magnetic cycles become weaker while still intermittently exhibiting large-scale dynamo action. This regime allows for the Rossby number (or equivalently, the dynamo number) to meander across the critical Rossby number for transition between different activity states or remain nearly constant, as hypothesized by \citet{metcalfe_tess_2020} for 94 Aqr Aa.  

To summarize, the {\it{subcritical solar magnetic dynamo}} hypothesis self-consistently explains the breakdown of stellar gyrochronology, the bimodal nature of the dynamo mechanism in the current Sun, solar cycle transition between grand minima and regular activity modes and the Sun's possible transition to a less magnetically active future.

\section*{Acknowledgments}
The authors thank Travis S. Metcalfe for insightful discussions. B.T. wishes to acknowledge the support extended to him during a Visiting Fellowship at the Center of Excellence in Space Sciences India (CESSI) at IISER Kolkata. CESSI is funded by the Ministry of Education, Government of India under the Frontier Areas of Science and Technology Scheme. 

\section*{Data Availability}
The data underlying this article will be shared on reasonable request to the corresponding author.

\bibliographystyle{mnras}
\bibliography{Subcritical_dynamo}

\end{document}